\documentstyle[epsf]{disk}
\begin{document}

\title{On accretion flow in Soft X-ray Transients}

\author{Piotr T.\ \.{Z}YCKI and Chris DONE\\
{\it Department of Physics, University of Durham,
South Road, Durham DH1 3LE, England, U.K.} \\
David A.\ SMITH \\
{\it Department of Physics and Astronomy, University of Leicester,
       University Road, Leicester LE1 7RH, England, U.K.}}

\maketitle

\section*{Abstract}

We have analyzed archival {\it Ginga\/} data of a number of Soft X-ray
Transient sources and modelled them in an attempt to constrain the geometry
and physical properties of accretion. Our models include a self-consistent
description of X-ray reprocessing, linking the properties of the reflected
continuum and those of the iron fluorescence K$\alpha$ line, and allowing
for relativistic smearing of spectral features.  We derive
constraints on the inner extent of the disk and its ionization,
assuming a generic geometry
of a central X-ray source and an external accretion disk.

Fitting our models to the data we have found that in some sources -- 
GS~1124-68 (Nova Muscae 1991) and
GS~2000+25 -- the evolution during decline proceeded in qualitative
agreement with recent theoretical models based on advection-dominated
solutions of accretion flow. These models link the high/soft--low/hard 
transition with a change of geometry, most importantly, the inner radius
of the optically thick disk. Quantitatively, our analysis requires serious 
revision of the model assumptions.  
A remarkably exceptional case is that of GS~2023+338,
where the source behavior both during outburst and decline was quite 
different than the other, more ``standard'' ones.

\section{Introduction}

Accretion flow around black holes in compact objects (both active galaxies
and stellar sources) clearly proceeds
through at least two phases: optically thick, cooler plasma producing a
soft thermal component, and optically thin, hot plasma where hard 
X-rays/$\gamma$-rays are produced (see Mushotzky et al.\ 1993 and
Tanaka \& Lewin 1995 for reviews).

The geometrical configuration of the two phases is still uncertain. Simple
possibilities include pure radial and pure vertical stratification of the
flow, but more complicated geometries, with partial overlap are also 
considered. Localized active regions above the disk, appearing due to 
e.g.\ magnetic activity, are
also considered (Galeev et al.\ 1979; Stern et al.\ 1995).

The commonly
made assumption that the observed hard X-ray spectra are due to the 
inverse Compton process requires an input of soft photons from the optically
thick plasma to the optically thin one. 
The reverse interaction must then also
take place: the hard X-rays will illuminate the optically thick plasma 
(Guilbert \& Rees 1988) and they will be reflected and reprocessed.
As a result, a continuum spectral component is formed by Compton 
reflection, with spectral features due to emission/absorption by heavy
elements (primarily iron) superimposed on it.

Studying the effects of such reprocessing can be 
extremely useful as a tool
for constraining the spatial distribution of the two phases of accreting 
plasma.
The amplitude of the reflected component constrains the solid angle subtended
by the optically thick plasma from the X-ray source. The shape of the reflected
continuum below $\sim 10$ keV, and the energy and strength of iron spectral
features around 7 keV give information about ionization state of the reflecting
plasma. Possible broadening/smearing of spectral features can constrain the
plasma velocity field if it can be distinguished from e.g.\ broadening
due to Comptonization.

Black hole Soft X-ray Transients (SXT) are particularly suited to this 
kind of analysis.
Those are a sub-class of X-ray binaries that undergo occasional, dramatic
outbursts.
During the subsequent decline phase the mass accretion rate, which is 
thought to 
be  the most important parameter, decreases roughly exponentially, with 
a typical e-folding time scale of $30-60$ days. During the decline they
go through a sequence of well defined states characterized by distinct
spectral and timing properties (review in Tanaka \& Shibazaki 1996).

\section{Data}

For our analysis we use archival {\it Ginga\/} data of a number of SXT.
The data combine broad band (1 to 20--30 keV) with moderate spectral
resolution (18\% at 6 keV), very good statistics due to large effective
area of the LAC detector ($\sim 4000\,{\rm cm^{-2}}$) and systematic 
uncertainties of the instrument below 1\%. Background subtraction can pose a 
problem for bright sources, due to the contamination of the background
monitors by source counts. We have developed a method to estimate the 
background, described in some detail in \.{Z}ycki et al.\ (1998b).

\section{Models}

\begin{figure}
\begin{center} \leavevmode
\hbox{%
\epsfysize=4cm 
\epsffile[60 250 578 664]{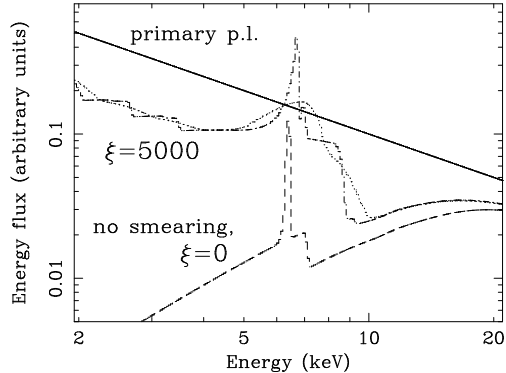}}
\end{center}
\caption{X--ray reprocessed spectra showing effects of ionization 
(dashed vs.\ dot-dashed curves) 
and relativistic smearing
(dotted vs.\ dot-dashed curves).
}
\label{zds:fig_reproc}
\end{figure}

Our model of X-ray reprocessing, described in details in \.{Z}ycki et
al.\ (1998b), self-consistently combines the model for Compton--reflected 
continuum and Monte Carlo computations of the iron K$\alpha$ line.
The main model parameter is the ionization parameter,
$\xi\equiv 4\pi F/n$, where $F$ is the illuminating flux and $n$ electron
density.
The amplitude of the component is expressed as a solid angle of the reflector,
normalized to $2\pi$, $f\equiv \Omega/2\pi$.
The effects of relativistic smearing are computed convolving the total
reprocessed spectrum with relativistic line profiles implemented in 
{\sc XSPEC} 
as the ``diskline'' model. 
The main parameter here is the inner disk radius in units of
$R_{\rm g} \equiv G M/c^2$, assuming a fixed illumination emissivity 
(we assume $F_{\rm irr}(r) \propto r^{-3}$). Figure~\ref{zds:fig_reproc} 
shows typical model spectra.

\section{Results}


\subsection{Low/hard state}

\begin{figure}
\begin{center} \leavevmode
\hbox{%
\epsfysize=5.5cm 
\epsffile[100 230 500 450]{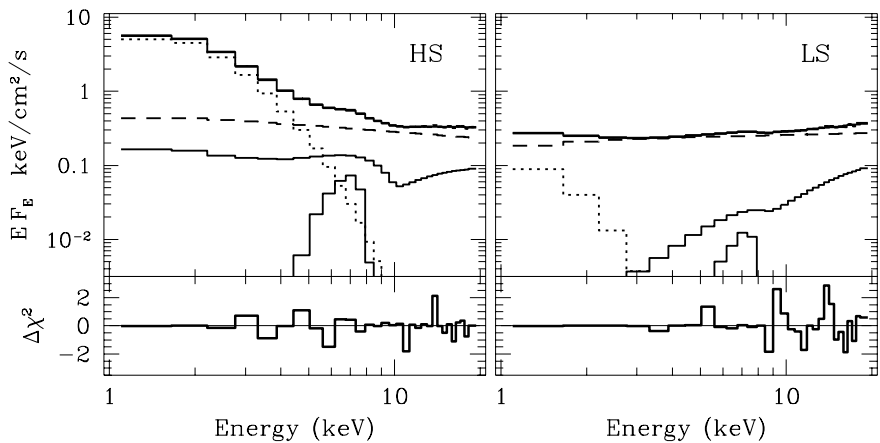}}
\end{center}
\caption{Spectra of Nova Muscae in low and high states (upper 
panels) and $\chi^2$ contributions (lower panels). Lower
solid histograms show the iron K$\alpha$ line and the reflected continuum,
dotted histograms show the soft thermal component and the dashed histograms
show the primary power law. See Table~\ref{zds:tab_fits} for fits details.
}
\label{zds:fig_HS_LS}
\end{figure}

The results are shown in Table~\ref{zds:tab_fits} The reprocessed component
is significantly present in the spectra. Its amplitude is significantly smaller
than 1, typical values being 0.3 -- 0.7. It is rather weakly ionized, 
$\xi < 50 $ corresponding to mean ionization level of iron $<$FeXVII
i.e.\ only M-shell electrons removed.
Relativistic smearing of the reprocessed component is significant in some
of the spectra, and is consistent with decreasing in time, as the sources 
progressed through the decline phase. Even at the beginning of decline 
in GS~2023+338 the smearing is less than expected from a disk extending 
down to the last
stable orbit at $r_{\rm ms} = 6\,R_{\rm g}$, typical values being 
$R_{\rm in} \sim 20-30
\,R_{\rm g}$ (see Figure~\ref{zds:fig_HS_LS} and \.{Z}ycki et al.\ 1997),
although $R_{\rm in}$ is compatible with $r_{\rm ms}$ in Nova Muscae
(\.{Z}ycki et al.\ 1998a).

\begin{table}[t]
\caption{Results of fitting the reprocessing model to the data} 
\label{zds:tab_fits}
\vspace{.5pc}
\begin{center}
\ixpt
\begin{tabular}{c|llllcll} \hline
sp. st. & Source &  
obs.\ date & 
~~~~~~$\Gamma$ & ~~~~~$f$ & $\xi$ & 
   $R_{\rm in}\ (R_{\rm g})$ &   $\chi^2$/dof \\
\hline
LS & GS 1124 & 
23/07/91 & 
$1.72\pm 0.02$  &  $0.24^{+0.11}_{-0.08}$  &  
     $17^{+40}_{-16}$  &$ 50^{+\infty}_{-35}  $ & 15.3/24 \\
LS & GS 2023 & 
19/06/89 & 
$1.72\pm 0.03$  &  $0.66^{+0.11}_{-0.07} $  &  
     $0^{+1} $  &
   $ 18^{+13}_{-7}$ & 23.1/23 \\
LS & GS 2000 & 
16/12/88 & 
$1.93^{+0.14}_{-0.08} $  &  $0.24^{+0.80}_{-0.21} $  &  
   $ 100^{+5\times 10^4}_{-100} $ & -- & 20/24  \\
\hline
HS & GS 1124 & 
18/05/91 & 
  $2.29^{+0.05}_{-0.03} $  &  
      $0.64^{+0.40}_{-0.10} $  &  $(3.5^{+9.5}_{-3.0})\times 10^4 $  &
                $ 18^{+22}_{-6} $ & 13.5/22 \\
HS & GS 2000 & 
 18/10/88 & 
$2.00^{+0.26}_{-0.03} $  &  $0.45^{+0.45}_{-0.11} $  &  
  $(15^{+45}_{-14.6})\times 10^3$ &
   $ 7^{+2}_{-1} $ & 15/21 \\
\hline
VHS & GS 1124 & 
11/01/91 &  
$2.02^{+0.16}_{-0.21} $  &  $0.35^{+0.25}_{-0.09} $  & 
 $(10^{+10}_{-8})\times 10^3 $  &
   $ 13^{+6}_{-3}  $ & 26.6/31 \\
VHS(?) & GS 2023 & 
30/05/89 & 
$1.70\pm 0.01 $  &  $0.17^{+0.16}_{-0.04} $  &  
 $(6^{+18}_{-4})\times 10^3 $  &   $ 6^{+2}  $ & 25.7/27 \\
\hline
\end{tabular}
\end{center}
\end{table}

\subsection{High/soft state}

In the high state spectra the major difference in the reprocessed component 
is its strong ionization (Table~\ref{zds:tab_fits}). Dominant iron ions are 
usually He- and H-like ones. Consequently, the K$\alpha$ line and 
absorption edges are much stronger in the composite spectra. The amplitude
$f$ is also larger, close to 1. Relativistic smearing is significant, but
the values of $R_{\rm in}$ differ between different objects. The fact that,
for GS~2000+25, $R_{\rm in} \approx 6$ (which is not expected given the 
assumed  $F_{\rm irr} \propto r^{-3}$ i.e.\ without the boundary condition 
term; Shakura \& Sunyaev 1973) 
may suggest that an additional broadening
of that component (e.g.\ due to Comptonization) takes place.

We find that the soft component of the continuum spectrum cannot be 
described by 
either pure black body or  multi-temperature (disk) blackbody spectrum. 
Additional
power law tail is required, which can be modelled as additional Comptonization
of a (disk) black body. This differs from results of Ebisawa et al.\ (1994)
who were able to fit the disk blackbody to the spectra but they were using
the smeared edge model with a narrow gaussian to model the iron spectral
features, rather than a reprocessing model.

\subsection{Very high state}

Nova Muscae showed an example of Very High State spectrum at the peak of its 
outburst (Jan--Feb 1991). 
On 11th Jan the soft continuum component again cannot be modelled as a 
(disk) blackbody,  so we assume a Comptonized blackbody instead 
(Figure~\ref{zds:fig_gs}, middle panel).
The presence of the reprocessed component is then again highly statistically
significant although its amplitude is not well constrained from above, in
particular if radial distribution of ionization is allowed for.
The reflection is strongly ionized and smeared, similarly to high state 
spectra (Table~\ref{zds:tab_fits}). Thus although the overall spectrum is
dominated by the hard power law ($\Gamma\sim 2$) as in LS, the properties
of both the soft component and the reprocessing are similar to HS, so the
suggested histeretic behavior of SXT (Miyamoto et al.\ 1995) does not seem
to be complete.

\section{The exceptional case of GS~2023+338}

\begin{figure}
\begin{center} \leavevmode
\hbox{%
\epsfxsize=0.97 \textwidth
\epsffile[25 480 570 670]{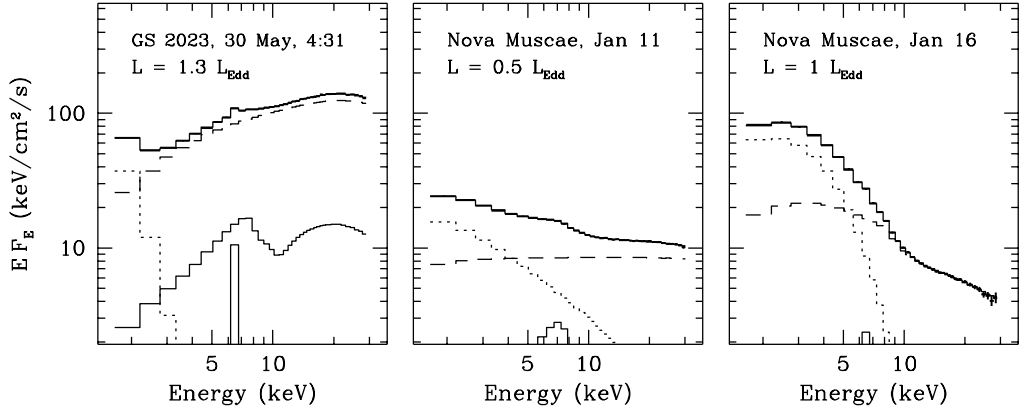}}
\end{center}
\caption{Comparison of two spectra of Nova Muscae in Very High State
and a spectrum of GS~2023+338 at the very peak of its outburst. The 
luminosities are bolometric ones but the correction is rather uncertain
in the case of GS~2023.
}
\label{zds:fig_gs}
\end{figure}

GS~2023+338 was clearly an unusual source, showing fast and chaotic
flux and spectral variability, and no clear soft, thermal component even
when its luminosity was very high (Tanaka \& Lewin 1995). 

The spectrum of GS~2023+338 at the peak of its outburst was rather unusual:
it can be described by an optically thick comptonization  of a disk blackbody 
radiation of rather high temperature, $k T_0 \approx 1.4\,$keV in a rather cool
plasma cloud, $k T_{\rm e}\approx 10\,$keV, $\tau \approx 6$.
Comparison with spectra of Nova Muscae at a similar luminosity level
(Figure~\ref{zds:fig_gs}) suggests that the actual mass accretion rate
in GS~2023 might have been super-Eddington. The hypothesis that the observed
emission is a (quasi-thermal) disk radiation is not however easily compatible
with the presence of spectral features due to highly ionized iron, since
in such a geometry disk irradiation is rather ineffective.

\begin{figure}
\begin{center} \leavevmode
\hbox{%
\epsfxsize=0.9 \textwidth
\epsffile[25 460 570 670]{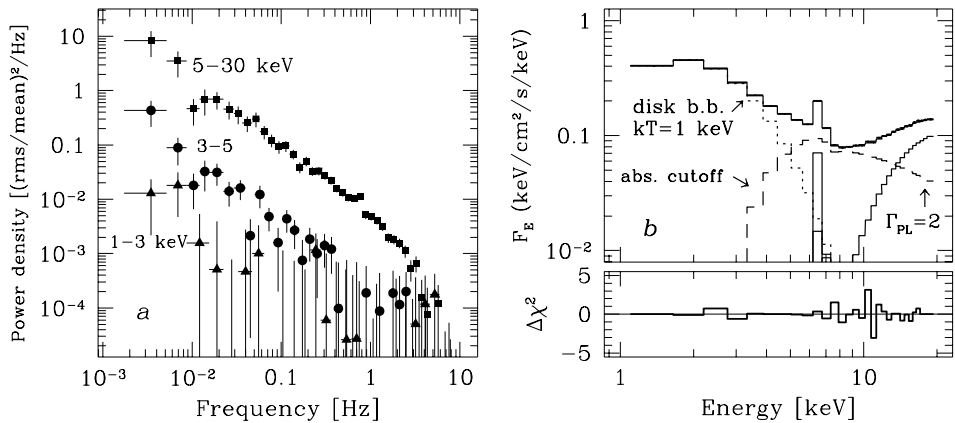}}
\end{center}
\caption{Power density spectra in three energy bands ({\it a\/}) and 
energy spectrum ({\it b\/})
of GS~2023+338 on 30th May (around 7:00 GMT), suggesting that the source was 
in the high/soft state, but it coincided with heavy absorption.
}
\label{zds:fig_soft}
\end{figure}

The short time-scale ($\sim$few sec) flux and spectral variability can be 
attributed to strong
and rapidly variable photo--electric absorption. During one the episodes
of such strong absorption on 30th May, the source spectrum is consistent
with a presence of a soft thermal component of temperature $T\approx 1\,$keV,
and a steep power law tail, $\Gamma \approx 2$. Such  spectra are typically
observed in high/soft states of GBH (Figure~\ref{zds:fig_soft}; \.{Z}ycki
et al.\ 1998c).

During its decline from outburst the source showed a fairly regular, 
exponential decline of the luminosity ($t_0\sim 30\,$ days) with again 
rapid variability due to photo-electric absorption superimposed on it. 
The amplitude of the reprocessed component decreases with time,
in rough correlation with decreasing level of relativistic smearing, 
i.e.\ the behavior is consistent with $R_{\rm in}$ increasing. However,
the power law spectral index remains practically constant,
in sharp contrast to the behavior of GS~1124-68, GS~2000+25 and GX~339
(\.{Z}ycki et al.\ 1998b).

\section{Discussion}

The results presented in previous Sections clearly suggest that the
geometry of accretion and the physical state of reprocessing material
change as sources make transition between hard and soft states.
The most significant changes to the reprocessing are those of 
ionization of the reflector and the amplitude of reflection.
In the low/hard state the solid angle of the reprocessor from the X-ray source 
is smaller and the reprocessor is colder than in the high/soft state. 
The amount
of relativistic smearing does not show clear, statistically significant 
changes between the transitions, but $R_{\rm in}$ is consistent with
increasing as the amplitude $f$ decreases.

We can compare the above results with predictions of models proposed for the
spectral transitions.

The observed changes are qualitatively  expected in the models of  
Mineshige (1996) and Esin et al.\ (1997). These models assume predominantly
radial stratification of the flow, with optically thick disk extending only
outside certain transition radius. The central, hot solution derives from
advection dominated accretion flow solutions although when applied to 
typical low state, it has radiative efficiency of $\sim 0.35$. 
The transition radius (which can be identified with our $R_{\rm in}$) has 
to decrease with increasing  $\dot m$ to explain softer spectra when 
sources are brighter,
and this also explains larger amplitude of the reprocessed component and its
strong ionization in high state. The behavior of $R_{\rm in}$ 
we find from
relativistic smearing is also consistent with that prediction.
Thermal radiation emitted by the disk can by Comptonized by hot upper layer
of the disk, giving rise to the observed steep power law tail.

Quantitatively, however, the values of $R_{\rm in}$ 
we derive are rather smaller than those postulated by Esin et al. 
In particular, in the low state we derive $R_{\rm in} \sim 20-50 \, R_{\rm g}$,
while they assume $2\times 10^4 \, R_{\rm g}$. Esin et al.\ (1998) argue
that  the smaller values are still compatible with their model but this claim
has not been verified by actual data modelling.

Relatively small values of $R_{\rm in}$ we derive indicate that there must
exist significant overlap between the central hot X-ray source and external 
cool disk i.e.\ the hot source has to extend much further out than 
$R_{\rm in}$: otherwise the energy  release within the source would be much
too small to account for the overall energetics and spectral shape 
(Gierli\'nski et al.\ 1997; review in Poutanen 1998). 
Perhaps the hot plasma forms a corona at large distances. The transition
between the disk and the corona at large distances could result from e.g.\ 
the thermal instability of X-ray irradiated plasma, as considered by 
Witt et al.\ (1997).

An alternative geometry was proposed by Di Matteo et al.\ (1998).
They suggest that magnetic flares above the disk produce the hard X-ray
radiation, the disk itself always extends to the last stable orbit, and the
hard/low -- soft/high transition is due to changes in characteristic height
of the flares. In order to account for harder spectra in the low state they 
postulate that the flares are higher above the disk in this state. 
This however does not 
naturally explain why the reflection amplitude is smaller in the hard state
unless the
flares are strongly concentrated towards the inner edge of the disk, so
a significant fraction of photons is lost within the central hole. 
An alternative explanation, that the inner
disk is strongly ionized and so the iron  spectral features are weak, does
not seem to be supported by data analysis, at least in Cyg X-1 (Done 
\& \.{Z}ycki 1998).

\section{References}

\re 
Di Matteo T. et al.\ 1998, MNRAS, submitted (astro-ph/9805345)
\re 
   Done C., \.{Z}ycki P. T.\ 1998, MNRAS, submitted
\re 
   Ebisawa K. et al.\ 1994, PASJ 46, 375
\re 
   Esin A. A., McClintock J. E., Narayan R.\ 1997, ApJ 489, 865
\re 
   Esin A. A., Narayan R., Cui W., Grove J. E., Zhang S.-N.\ 
1998, ApJ 505, 854
\re 
   Galeev A. A., Rosner R., Vaiana G. S.\ 1979, ApJ 229, 318
\re 
   Gierli\'{n}ski M. et al.\ 
1997, MNRAS 288, 958
\re 
   Guilbert P. W., Rees M. J.\ 1988, MNRAS 233, 475
\re
  Mineshige S.\ 1996, PASJ 48, 93
\re
  Miyamoto S. et al.\ 1995, ApJ 442, L13
\re
   Mushotzky R. F., Done C., Pounds K. A.\ 1993, ARA\&A 31, 717
\re 
   Poutanen J.\ 1998, in Theory of Black Hole Accretion Discs,
   eds.\ M.\ A.\ Abramowicz, G.\ Bj\"{o}rnsson, J.\ E.\ Pringle 
  (CUP, Cambridge) (astro-ph/9805025)
\re 
   Shakura N. I., Sunyaev R. A.\ 1973, A\&A 24, 337
\re
   Stern B. E. et al.\ 1995, ApJ 449, L13
\re 
   Tanaka Y.,  Lewin W. H. G.\  1995, in  X--Ray Binaries, eds.\  W.\ H.\ G.\ 
   Lewin, J.\ van Paradijs, E.\ van den Heuvel, (CUP, Cambridge) 
\re 
 Tanaka Y., Shibazaki N.\ 1996, ARA\&A 34, 607
\re    
  Witt H. J., Czerny B., \.{Z}ycki P. T.\ 1997, MNRAS 288, 848
\re  
 \.{Z}ycki P. T., Done C.,  Smith D. A.\ 1997, ApJ 488, L113 
\re  
 \.{Z}ycki P. T., Done C.,  Smith D. A.\ 1998a, ApJ 496, L25 
\re  
 \.{Z}ycki P. T., Done C.,  Smith D. A.\ 1998b, MNRAS, submitted 
    (astro-ph/9811106)
\re  
\.{Z}ycki P. T., Done C.,  Smith D. A.\ 1998c, in preparation

\end{document}